\documentclass[12pt]{article}

\usepackage{color}
\usepackage{a4}
\usepackage{latexsym}
\usepackage{cite}
\usepackage{lineno}

\usepackage{array}
\newcolumntype{T}{>{\ttfamily} c}
\newcolumntype{M}{>{$\displaystyle} c <{$}}

\usepackage{epsfig}
\usepackage{graphicx}

\usepackage{pslatex}
\usepackage[latin1]{inputenc}
\usepackage[T1]{fontenc}

\usepackage{color,colordvi}
\def\colour4colour#1{\Blue{#1}}
\newcommand{\colourcolour}[1]{{\color{blue}{#1}}}

\newcommand{\lsim}{\raisebox{-0.7mm}{$\:\:\stackrel{<}{{\scriptstyle
 \sim}}\:\: $} }

\def\slash#1{\rlap{\hbox{$\mskip 1 mu /$}}#1}      % good slash for lower case
\newcommand{\beq}{\begin{equation}}
\newcommand{\eeq}{\end{equation}}
\newcommand{\bea}{\begin{eqnarray}}
\newcommand{\eea}{\end{eqnarray}}
\newcommand{\nn}{\nonumber}

\newcommand{\ra}{\rightarrow}
\newcommand{\equal}{\:\: = \:\:}

\newcommand{\als}{\alpha_{\rm s}}
\newcommand{\ars}{a_{\rm s}}

\newcommand{\hspn}{{\hspace{-5mm}}}
\newcommand{\hspp}{{\hspace{5mm}}}

\def\frct#1#2{\mbox{\small{$\displaystyle\frac{#1}{#2}$}}}

\def\as(#1){{\alpha_{\rm s}^{\,#1}}}
\def\ar(#1){{a_{\rm s}^{\,#1}}}

\def\zr#1{{\zeta_{\:\!#1}^{}}}
\def\mus{{\mu^{\,2}}}

\def\B(#1,#2){{\beta_{#1}^{\,#2}}}

\def\nc{{n_c}}
\def\ncs{{n_{c}^{\,2}}}
\def\nct{{n_{c}^{\,3}}}
\def\ncf{{n_{c}^{\,4}}}
\def\camcf{(\ca-\cf)}
\def\ca{{C^{}_A}}
\def\cas{{C^{\,2}_A}}

\def\cf{{C^{}_F}}
\def\cfs{{C^{\, 2}_F}}
\def\cft{{C^{\, 3}_F}}

\def\nf{{n^{}_{\! f}}}

\def\nfo{{n^{\,1}_{\! f}}}
\def\nfs{{n^{\,2}_{\! f}}}
\def\nft{{n^{\,3}_{\! f}}}

\def\dfRRnr{{ {d_{\,R}^{\,abcd} d_{\,R}^{\,abcd} \over n_c} }}
\def\dfRRnR{{ {d_{\,R}^{\,abcd} d_{\,R}^{\,abcd} / n_c} }}

\def\DNn#1{D_0^{\:#1}}
\def\DNm#1{D_{-1}^{\:#1}}
\def\DNp#1{D_1^{\:#1}}
\def\DNpp#1{D_2^{\:#1}}

\def\xm1{{(1 \! - \! x)}}
\def\xp1{{(1 \! + \! x)}}

\def\Lnt(#1){\ln^{\,#1}(1\!-\!x)}
\def\pqq(#1){p_{\rm{qq}}(#1)}

\def\S(#1){{{S}_{#1}}}
\def\Ss(#1,#2){{{S}_{#1,#2}}}
\def\Sss(#1,#2,#3){{{S}_{#1,#2,#3}}}
\def\Ssss(#1,#2,#3,#4){{{S}_{#1,#2,#3,#4}}}
\def\Sssss(#1,#2,#3,#4,#5){{{S}_{#1,#2,#3,#4,#5}}}
\def\Ssssss(#1,#2,#3,#4,#5,#6){{{S}_{#1,#2,#3,#4,#5,#6}}}
\def\Sssssss(#1,#2,#3,#4,#5,#6,#7){{{S}_{#1,#2,#3,#4,#5,#6,#7}}}

\def\Sp(#1,#2){{{S}_{#1}^{\,#2}}}

\def\H(#1){{\rm{H}}_{#1}}
\def\Hh(#1,#2){{\rm{H}}_{#1,#2}}
\def\Hhh(#1,#2,#3){{\rm{H}}_{#1,#2,#3}}
\def\Hhhh(#1,#2,#3,#4){{\rm{H}}_{#1,#2,#3,#4}}
\def\Hhhhh(#1,#2,#3,#4,#5){{\rm{H}}_{#1,#2,#3,#4,#5}}
\def\Hhhhhh(#1,#2,#3,#4,#5,#6){{\rm{H}}_{#1,#2,#3,#4,#5,#6}}

\begin{document}
\setlength{\parskip}{0.2cm}
\setlength{\baselineskip}{0.54cm}

% --------------------------------------------------------------------

\begin{titlepage}
\noindent
DESY 23--022 \hfill February 2023\\
LTH 1333 \\
\vspace{0.6cm}
\begin{center}
{\LARGE \bf Four-loop splitting functions in QCD \\[1ex]
  -- The quark-quark case --}\\ 
\vspace{1.6cm}
\large
G.~Falcioni$^{\, a}$, F.~Herzog$^{\, a}$, S. Moch$^{\, b}$ 
and A. Vogt$^{\, c}$\\

\vspace{1.2cm}
\normalsize
{\it $^a$Higgs Centre for Theoretical Physics, School of Physics and Astronomy\\
  The University of Edinburgh, Edinburgh EH9 3FD, Scotland, UK}\\
\vspace{4mm}
\vspace{4mm}
\normalsize
{\it $^b$II.~Institute for Theoretical Physics, Hamburg University\\
\vspace{0.5mm}
Luruper Chaussee 149, D-22761 Hamburg, Germany}\\
\vspace{4mm}
\vspace{4mm}
{\it $^c$Department of Mathematical Sciences, University of Liverpool\\
\vspace{0.5mm}
Liverpool L69 3BX, United Kingdom}\\
\vspace{3cm}
{\large \bf Abstract}
\vspace{-0.2cm}
\end{center}
We have computed the even-$N$ moments $N\leq 20$ of the pure-singlet 
quark splitting function $P_{\,\rm ps}$ at the fourth order of 
perturbative QCD via the anomalous dimensions of off-shell flavour-singlet 
operator matrix elements.
Our results, derived analytically for a general gauge group, agree with all 
results obtained for this function so far, in particular with the lowest six 
even moments obtained via physical cross sections.
Using these results and all available endpoint constraints, we construct
approximations for $P_{\rm ps}$ at four loops that should be sufficient for 
most collider-physics applications.
Together with the known results for the non-singlet splitting function 
$P_{\rm ns}^{\,+}$ at this order, this effectively completes the quark-quark 
contribution for the evolution of parton distribution at N$^{\:\!3}$LO 
accuracy.
Our new results thus provide a major step towards fully consistent 
N$^{\:\!3}$LO calculations at the LHC and the reduction of the residual
uncertainty in the parton evolution to the percent level. 
\vspace*{0.5cm}
\end{titlepage}

\newpage
% --------------------------------------------------------------------

Collinear factorization in Quantum Chromodynamics (QCD) is controlled by 
a set of universal functions, the splitting functions $P$, governing 
the scale dependence of the parton distribution functions (PDFs), 
which are essential ingredients in all theoretical predictions for 
scattering processes with initial-state hadrons~\cite{Accardi:2016ndt}.
The splitting functions are calculable in the perturbative approach to 
QCD. They have been known, for a long time, at three-loop accuracy 
\cite{Moch:2004pa,Vogt:2004mw}, which is the next-to-next-to-leading 
order (NNLO or N$^2$LO) in the expansion in powers of the strong
coupling $\als$.
 
This theoretical accuracy, however, faces challenges due to the precision
of the currently available experimental data collected at the 
Large Hadron Collider (LHC). Also, the expectations for the LHC's Run 3, 
the high-luminosity phase (HL-LHC) as well as the plans for the future 
Electron-Ion Collider (EIC)~\cite{AbdulKhalek:2021gbh} indicate the need 
to increase the precision by one quantum loop to N$^3$LO accuracy. 
This will reduce the uncertainty in the parton evolution to percent-level
precision. 
It is also required for a fully consistent use of the available N$^3$LO 
QCD predictions for hard scattering cross sections for key processes 
\cite{Anastasiou:2015vya,Duhr:2020seh,Chen:2021isd,Baglio:2022wzu}
in proton-proton collisions at at the LHC as well as for structure 
functions in deep-inelastic electron-proton scattering (DIS) probed by  
the EIC \cite{Vermaseren:2005qc,Moch:2008fj,Davies:2016ruz}.

The PDFs of quarks and antiquarks of flavour $i$ and of the gluon,
$q_i^{}(x,\mus)$, $\bar{q}_i^{}(x,\mus)$ and $g(x,\mus)$, are given by 
the respective number distributions in the fractional hadron momentum 
$x$ at the factorization scale $\mu$. 
For $\nf$ flavours, the flavour-summed quark distributions
\beq
\label{eq:sgPDFs}
  q_{\rm s}^{}(x,\mus) \; = \; \sum_{i=1}^{\nf} 
  \left[\,q_i^{}(x,\mus) + \,\bar{q}_i^{}(x,\mus) \:\!\right] 
\eeq
define the singlet quark distribution $q_{\rm s}^{}$, which mixes under 
QCD evolution with the gluon distribution $g$ through the matrix 
of splitting functions $P_{\,\rm ij}$ as 
\beq
\label{eq:sgEvol}
  \frac{d}{d \ln\mus} \; 
  \Big( \begin{array}{c} \! q_{\rm s}^{} \!\! \\ \!g\!  \end{array} \Big)
  \: = \: \left( 
  \begin{array}{cc} \! P_{\rm qq} & P_{\rm qg} \!\!\! \\
                    \! P_{\rm gq} & P_{\rm gg} \!\!\! \end{array} \right) 
  \otimes
  \Big( \begin{array}{c} \!q_{\rm s}^{}\!\! \\ \!g\!  \end{array} \Big)
\eeq
where $\otimes$ stands for the Mellin convolution in the momentum 
variable $x$. 
The perturbative expansion in the strong coupling with 
$\ars \:\equiv\; \als(\mus)/(4\pi)$ can be written as
\beq
\label{eq:Pexp}
  P_{\,\rm ij}^{}\left(x,\als\right) 
  \; = \; \sum_{n=0}\, \ar(n+1) P^{\,(n)}_{\,\rm ij}(x)
  \:\: ,
\eeq
such that the terms $P^{\,(k)}_{\,\rm ij}$ correspond to the QCD 
predictions at N$^k$LO accuracy. Here and below we identify, without 
loss of information, the renormalization scale with the scale $\mu$
in eq.~(\ref{eq:sgPDFs}).

The quark-quark splitting function $P_{\rm qq}$ in eq.~(\ref{eq:sgEvol}) 
is the sum $P_{\rm qq} = P_{\rm ns}^{\,+} + P_{\rm ps}^{}$, 
where $P_{\rm ns}^{\,+}$ denotes the splitting function for non-singlet
combinations of the quark-antiquark sums $q_i^{}+\bar{q}_i^{}$,
and $P_{\,\rm ps}$ is the pure-singlet contribution.
The results for the four-loop (N$^3$LO) non-singlet splitting functions 
$P^{\,(3)}_{\,\rm ns}$ are complete in the planar limit, i.e., the 
large-$\nc$ limit of a colour SU$(\nc)$ gauge group, 
and very good approximations for the remaining non-planar terms have 
been obtained \cite{Moch:2017uml,MVV-tba}. 

In this letter, we address the computation of the pure-singlet part 
$P^{\,(3)}_{\,\rm ps}$ at N$^3$LO.
Together with the result for $P_{\,\rm ns}^{(3)+}$, this provides the 
quark-quark splitting functions at four loops.
We present analytic results for the first 10 even Mellin moments 
$N \leq 20$ of $P^{\,(3)}_{\,\rm ps}(x)$, valid for a general compact 
simple gauge group.
These correspond (up to a negative sign, which is a standard convention) 
to the anomalous dimensions $\gamma^{\,(3)}_{\,\rm ps}$,
\beq
\label{eq:gamP}
  \gamma_{\,\rm ij}^{\,(n)}(N) \; = \; - \int_0^1 \!dx\:\, x^{\,N-1}\, 
  P_{\,\rm ij}^{\,(n)}(x) 
\:\: .
\eeq
This extends previous results on this function, which is already known 
in the limit of large numbers of flavours $n_f$~\cite{Davies:2016jie}.
Following the approach of refs.~\cite{Moch:2004pa,Vogt:2004mw} via 
physical quantities in inclusive DIS, also low moments 
($N \leq 8$) of $P^{\,(3)}_{\rm ps}$ have been presented before 
\cite{Moch:2021qrk}, which have been used in approximations 
\cite{McGowan:2022nag} for applications in N$^{3}$LO PDF fits.
For the parts proportional to the quartic colour factor 
$d_R^{\,abcd}d_R^{\,abcd}$ (see below), the moments up to $N = 16$ 
have been derived~\cite{Moch:2018wjh} by computing anomalous 
dimensions of off-shell flavour-singlet operator matrix elements (OMEs).

Here we extend the method of OMEs further. 
The starting point is the standard set of the spin-$N$ twist-two 
irreducible flavour-singlet quark and gluon operators, given by
\bea
\label{eq:loc-ops}
 O_{\rm q}^{\{\mu^{\,}_1,...,\mu^{\,}_N\}} 
 & \!=\! & \frct{1}{2}\,\overline{\psi}\, 
 \gamma^{\,\{\mu^{}_1}D^{\,\mu^{}_2}\ldots D^{\,\mu^{}_N\}}\,\psi 
 \,+\, \mbox{traceless}
 \: , 
\nn \\
 O_{\rm g}^{\{\mu^{\,}_1,...,\mu^{\,}_N\}} 
 & \!=\! & \frct{1}{2}\,F^{\nu \{ \mu^{}_1} 
   D^{\,\mu^{}_2}\cdots D^{\,\mu_{N-1}}\, F^{\mu^{}_N \}}_{\hspace*{4mm} 
   \nu} \,+\, \mbox{traceless}
 \: ,
\eea
where $\psi$ represents the quark field, $F_{\mu\nu}$ the gluon 
field-strength tensor and $D_{\,\mu}=\partial_{\,\mu}-{\rm i}gA_{\,\mu}$ 
the covariant derivative with the coupling $g$, where 
$g^2/(4\pi) = \als$. 
The curly brackets $\{ \dots \}$ denote symmetrization.
Flavour non-singlet operators have been discussed in 
ref.~\cite{Moch:2017uml}.

Contraction of the Lorentz indices with $N$ identical light-like 
($\Delta.\Delta=0$) vectors $\Delta^{\,\mu}$ allows for a compact 
notation in terms of the quantities
\bea
  F^{\mu;a} \,=\, \Delta_{\,\nu}\,F^{\mu\nu;a}\, ,\qquad  
  A^a       \,=\, \Delta_{\,\mu} \, A^{\mu;a}\, ,\qquad
  D         \,=\, \Delta_{\,\mu}\,D^{\,\mu}\, ,\qquad 
  \partial  \,=\, \Delta_{\,\mu}\,\partial^{\,\mu}
  \, ,
\eea
with labels $a$ in the adjoint representation of the colour gauge group. 
The projection of eq.~(\ref{eq:loc-ops}) defines the scalar operators 
for even $N$ (suppressing the index $N$ here and below) as 
\bea
\label{eq:loc-ops-Delta}
 O_{\rm q} \, = \, 
 \frct{1}{2}\, \overline{\psi}\, \slash{\Delta}\,D^{\,N-2}\, \psi
 \: , \qquad
 O_{\rm g} \, = \, 
 \frct{1}{2}\, F_{\nu}^{\hspace*{1mm} a} \,D_{ab}^{\,N-2}\, F^{\nu;b} 
 \, .
\eea
These physical (gauge-invariant) operators, when evaluated in general Green's 
functions, mix under renormalization with non-physical, so-called alien 
operators, which also involve (anti-)ghost fields ${\overline c}$ and $c$. 
The general theory of the renormalization of gauge-invariant operators
has been worked out in a series of classical papers by Dixon and Taylor 
\cite{Dixon:1974ss}, Kluberg-Stern and \mbox{Zuber} 
\cite{Kluberg-Stern:1974nmx,Kluberg-Stern:1975ebk} and Joglekar and Lee 
\cite{Joglekar:1975nu,Joglekar:1976eb,Joglekar:1976pe}. 
These led the way to explicit computations at two loops in 
ref.~\cite{Hamberg:1991qt}, which solved an issue that had beset the 
pioneering calculation of ref.~\cite{Floratos:1978ny}, and 
refs.~\cite{Matiounine:1998ky,Blumlein:2022ndg}. 
Recently, the complete three-loop renormalization via a direct calculation 
of the alien counter-terms has been published~\cite{Gehrmann:2023ksf}. 

A general procedure to construct the basis of alien operators was formulated
by two of us  in ref.~\cite{Falcioni:2022fdm} and was used to construct an 
explicit basis up to four loops for any fixed spin $N$. 
Building on this work, we here consider the four-loop renormalization of 
singlet quark operators.
For this we need two sets of alien operators,  
$O_{A}^{\,i}=O_{\rm q}^{\,i}+O_{\rm g}^{\,i}+O_{\rm c}^{\,i}$ with $i=I,II$, 
which read 
\bea
\label{eq:alienI}
  O_{\rm q}^{\,I} &\!=\!& 
  \eta\, g\, \overline{\psi}\,\slash{\Delta}\, t^a\, \psi\, 
  \left(\partial^{\,N-2}A_a \right)
\, , \nn \\
  O_{\rm g}^{\,I} &\!=\!& 
  \eta\, \left(D.F\right)^a \left(\partial^{\,N-2}A_a\right) 
\, ,\nn \\
  O_{\rm c}^{\,I} &\!=\!& 
  \!\! -\eta\, \left(\partial \,\overline{c}^{\,a}\right) 
  \left(\partial^{\,N-1}c_a \right)
\eea
with a coupling $\eta$ which is a function of $N$ and $\als$, and 
\bea
\label{eq:alienII}
  O_q^{\,II} &\!=\!& 
  g^2\, \overline{\psi}\,\slash{\Delta}\, t_a\, \psi 
  \!\!\sum_{i+j=N-3}\, \kappa_{ij}^{} f^{\,abc}\, 
  \left(\partial^{\,i}A_b\right) \left(\partial^{\,j}A_c\right)  
\, ,\nn\\
  O_g^{\,II} &\!=\!& 
  g \left(D.F\right)_a \!\!\sum_{i+j=N-3}\!\! \kappa_{ij}^{}\, f^{\,abc}\, 
  \left(\partial^{\,i}A_b\right) \left(\partial^{\,j}A_c\right) 
\, , \nn\\
  O_c^{\,II} &\!=\!& 
  - g \!\!\sum_{i+j=N-3} \!\! \eta_{ij}^{}\, f^{\,abc}\, 
  \left(\partial \,\overline{c}_a\right) \left(\partial^{\,i}A_b\right) 
  \left(\partial^{\,j+1} c\right) 
\, , 
\eea
where $t^a$ ($f^{\,abc}$) denote the fundamental (adjoint) colour-group 
generators.
The couplings $\eta_{ij}^{}$ and $\kappa_{ij}^{}$ obey constraints due to 
the (anti-)BRST symmetry of the alien operators~\cite{Falcioni:2022fdm}, 
viz
\beq
  \label{eq:antiBRSTrel1}
  \eta_{ij}^{} \;=\;
  -\sum_{s=0}^{i}(-1)^{s+j}\,\Big(
  \begin{array}{c}
    \! s+j \! \\
    s
  \end{array}
  \Big)\,\eta_{(i-s)(j+s)}
  \,=\,
  2\, \kappa_{ij} + \eta \Big(
  \begin{array}{c}
    \! i+j+1 \! \\
    i
  \end{array}
  \Big)\, 
  \, .
\eeq
Hence the $\kappa_{ij}^{}$ are dependent on the $\eta_{ij}^{}$ for which we
obtain a compact expression in terms of binomial coefficients
\beq
\label{eq:etasol}
\eta_{ij}^{} \;=\; \eta\, \Big[(-1)^i- 3\,\Big(
  \begin{array}{c}
    \!N-2\! \\
    i
  \end{array}
  \Big)-\Big(
  \begin{array}{c}
    \! N-2\! \\
    i+1
  \end{array}
  \Big)  \Big]\,, 
\eeq
where, in the present case, the coupling $\eta=\eta(\als,N)$ from 
eq.~(\ref{eq:alienI}) is found to factorize to the loop-order required.

The physical operators in eq.~(\ref{eq:loc-ops-Delta}) mix under 
renormalization with the alien ones in eqs.~(\ref{eq:alienI}) and 
(\ref{eq:alienII}).
The latter are summarized from now on collectively as $O_{\!\rm A}$ 
and we denote renormalized operators as $[O]_{\rm i}^{}$, so that
\beq
\label{eq:renormmix}
\left( \begin{array}{c} 
\!\! O_{\rm q} \!\! \\ 
\!\! O_{\rm g} \!\! \\ 
\!\! O_{\rm A} \!\!    
\end{array} 
\right)
  \: = \: \left( 
  \begin{array}{ccc} 
    \!\! Z_{\rm qq} & Z_{\rm qg} & Z_{\rm qA} \!\! \\
    \!\! Z_{\rm gq} & Z_{\rm gg} & Z_{\rm gA} \!\! \\
    \!\! Z_{\rm Aq} & Z_{\rm Ag} & Z_{\rm AA} \!\! 
  \end{array} \right) 
\left( \begin{array}{c} 
\!\! $[$O$]$_{\rm q} \!\! \\  
\!\! $[$O$]$_{\rm g} \!\! \\ 
\!\! $[$O$]$_{\rm A} \!\!    
\end{array} 
\right)
  \:\: ,
\eeq
where the $Z$-factors are determined in terms of the anomalous dimensions 
$\gamma_{\,\rm ij}^{}$ in eq.~(\ref{eq:gamP}) (and the corresponding ones 
$\gamma_{\,\rm qA}$, $\gamma_{\,\rm gA}$ etc. including the alien operators)
through the standard renormalization group equation
\beq
\label{eq:Zfactors}
  \mu^{\:\!2}\frac{d}{d\mu^{\:\!2}}\,Z_{\,\rm ij} 
  \: = \,
\left(
\beta(\als)\, \frac{\partial}{\partial \als}
+ \gamma_{\,3}^{}\, \xi\, \frac{\partial}{\partial \xi} 
\right)\,Z_{\,\rm ij} 
  \: = \:
-\gamma_{\,\rm ik}\, Z_{\,\rm kj} 
  \, .
\eeq
Here $\beta$ is the QCD $\beta$-function and $\gamma_{\:\!3}^{}$ the gluon 
anomalous dimension, all known to a more than sufficiently high order 
\cite{Baikov:2016tgj,Herzog:2017ohr,Luthe:2017ttg,Chetyrkin:2017bjc}.
The $Z_{\rm ij}$ involving the alien operators can be gauge dependent, 
$\xi$ is the gauge parameter with $\xi=1$ for Feynman gauge.
Moreover, since the alien operators cannot mix into the set of physical
operators~\cite{Joglekar:1975nu,Joglekar:1976eb,Joglekar:1976pe}, the entries 
$Z_{\rm Aq}$ and $Z_{\rm Ag}$ in eq.~(\ref{eq:renormmix}) have to vanish.

\pagebreak

The setup of our OME computations follows previous 
work~\cite{Moch:2017uml,Moch:2018wjh}.
The necessary Feynman rules are determined from 
eqs.~(\ref{eq:loc-ops-Delta}) -- (\ref{eq:etasol}), which are sufficient 
for the present computations.
The diagrams for the OMEs 
$A_{\,\rm ij}\,=\, \langle\, {\rm j}(p) |\, O_{\rm i} \,|\, {\rm j}(p) \rangle$
with (physical or alien) spin-$N$ twist-two operators $O_{\:\!\rm i}$ inserted 
in Green's functions with off-shell quarks, gluons or ghosts 
have been generated using {\sc Qgraf}~\cite{Nogueira:1991ex}
and then processed, see ref.~\cite{Herzog:2016qas}, by a 
{\sc Form}~\cite{Vermaseren:2000nd,Kuipers:2012rf,Ruijl:2017dtg} 
program which collects self-energy insertions, determines the colour 
factors~\cite{vanRitbergen:1998pn} and classifies the topologies according 
to the conventions of the {\sc Forcer} package~\cite{Ruijl:2017cxj}. 
For computational efficiency, diagrams with the same colour factor and 
topology are merged into meta-diagrams. 

An optimized in-house version of {\sc Forcer}, briefly discussed in 
ref.~\cite{MVV-tba}, is employed to \mbox{perform} the integral reductions 
for fixed integer values of $N$. 
In practice, the range in $N$ is limited by the occurrence of high powers of 
scalar products in the loop integrals for high values of $N$, 
which lead to large-size expressions and long computing times in the 
topology transformations and parametric reductions encoded in {\sc Forcer}.
The divergences in the OMEs $A_{\rm ij}$ are treated in dimensional 
regularization with $D=4-2\epsilon$ dimensions, hence the $Z$-factors in 
eq.~(\ref{eq:renormmix}) are simple Laurent series in $\epsilon$ and the 
anomalous dimensions $\gamma_{\,\rm ij}$ can be read off from their single 
poles $1/\epsilon$.

For the quark-quark splitting functions, the physical OMEs $A_{\rm qq}$ 
have been obtained at even $N \leq 20$ up to four loops.
This includes both the flavour non-singlet parts \cite{MVV-tba} and the 
pure-singlet contributions addressed in the present letter.
The physical OMEs $A_{\rm qg}$, $A_{\rm gq}$ and $A_{\rm gg}$ and those
with the alien operators inserted into a quark two-point function, 
$A_{\rm Aq}$, have been computed up to three loops.
All others were needed at two loops only for the extraction 
of $P^{\,(3)}_{\,\rm ps}$ at four loops using eq.~(\ref{eq:Zfactors}). 
This leads to the following results for the N$^3$LO contributions to the
pure-singlet anomalous dimensions in eq.~(\ref{eq:gamP}) for QCD, i.e.,
the gauge group SU$(\nc=3)$,
\bea
\label{eq:gps3-numerics}
  \gamma_{\,\rm ps}^{\,(3)}(N\!=\!2) \; & =\! & 
  - 691.5937093 \,\nf
  + 84.77398149 \,\nfs
  + 4.466956849 \,\nft
\, , \nn \\
  \gamma_{\,\rm ps}^{\,(3)}(N\!=\!4) \; & =\! & 
  - 109.3302335 \,\nf
  + 8.776885259 \,\nfs
  + 0.306077137 \,\nft
\, , \nn \\
  \gamma_{\,\rm ps}^{\,(3)}(N\!=\!6) \; & =\! & 
  - 46.03061374 \,\nf
  + 4.744075766 \,\nfs
  + 0.042548957 \,\nft
\, , \nn \\
  \gamma_{\,\rm ps}^{\,(3)}(N\!=\!8) \; & =\! & 
  - 24.01455020 \,\nf
  + 3.235193483 \,\nfs
  - 0.007889256 \,\nft
\, , \nn \\
  \gamma_{\,\rm ps}^{\,(3)}(N\!=\!10) & =\! & 
  - 13.73039387 \,\nf
  + 2.375018759 \,\nfs
  - 0.021029241 \,\nft
\, , \nn \\
  \gamma_{\,\rm ps}^{\,(3)}(N\!=\!12) & =\! & 
  - 8.152592251 \,\nf
  + 1.819958178 \,\nfs
  - 0.024330231 \,\nft
\, , \nn \\
  \gamma_{\,\rm ps}^{\,(3)}(N\!=\!14) & =\! & 
  - 4.840447180 \,\nf
  + 1.438327380 \,\nfs
  - 0.024479943 \,\nft
\, , \nn \\
  \gamma_{\,\rm ps}^{\,(3)}(N\!=\!16) & =\! & 
  - 2.751136330 \,\nf
  + 1.164299642 \,\nfs
  - 0.023546009 \,\nft
\, , \nn \\
  \gamma_{\,\rm ps}^{\,(3)}(N\!=\!18) & =\! & 
  - 1.375969240 \,\nf
  + 0.960873318 \,\nfs
  - 0.022264393 \,\nft
\, , \nn \\
  \gamma_{\,\rm ps}^{\,(3)}(N\!=\!20) & =\! & 
  - 0.442681568 \,\nf
  + 0.805745333 \,\nfs
  - 0.020918264 \,\nft
\;.
\eea
The results for $N \leq 8$ agree with those obtained via cross sections for
inclusive DIS in ref.~\cite{Moch:2021qrk}.
As a further check, we have extended those DIS computations of 
$P_{\,\rm ps}^{\,(3)}$ to $N = 10$ and $N = 12$, their results also agree
with eq.~(\ref{eq:gps3-numerics}).
The large-$n_f$ parts agree with all-$N$ results of ref.~\cite{Davies:2016jie}.
In addition, the renormalization constants involving alien operators agree 
to the loop order required here with those recently published in 
ref.~\cite{Gehrmann:2023ksf}.

The analytic expressions for $\gamma_{\,\rm ps}^{\,(3)}$ for a general gauge 
group are given in app.~\ref{sec:appA} 
in eqs.~(\ref{eq:GpsN2}) -- (\ref{eq:GpsN20}).
They contain rational numbers and Riemann-$\zeta$ values, i.e. $\zeta_n$ with 
$n=3,4,5$. New all-$N$ results for $\gamma_{\,\rm ps}^{\,(3)}$ based on 
eqs.~(\ref{eq:GpsN2}) -- (\ref{eq:GpsN20}) and, in cases, even higher fixed-$N$ 
OME computations, have been derived for specific colour factors or terms 
proportional to certain Riemann-$\zeta$ values. 
They can be expressed in terms harmonic 
sums~\cite{Vermaseren:1998uu,Blumlein:1998if}, which are recursively defined by
\beq
\label{eq:Hsum}
  S_{\pm m_1^{},\,m_2^{},\,\ldots,\,m_d}(N) \:\:=\; \sum_{i=1}^{N}\:
  \frac{(\pm 1)^{i}}{i^{\, m_1^{}}}\: S_{m_2^{},\,\ldots,\,m_d}(i) 
\:\: ,
\eeq
and their weight $w$ is the sum of the absolute values of the indices~$m_d$.
In the results for the $n$-loop anomalous dimensions, quantities up to 
$w = 2\:\!n-1$ occur, which can be composed of harmonic sums,
Riemann-$\zeta_m$ values ($m \geq 3$), and simple denominators, 
$D_a^{\,k} = (N+a)^{-k}$. 
The latter count as objects of weight $m$ and $k$, respectively.
Due to (partial) conformal symmetry, especially the sums with the highest 
weights often arise with two specific (reciprocity respecting, cf.~the 
discussion in ref.~\cite{Moch:2017uml}) combinations of simple denominators,
\bea
\label{eq:etaDef}
  \eta &\!\equiv\!& \frct{1}{N} - \frct{1}{N+1} 
       \;\equiv\; \DNn{} - \DNp{}
       \;=\; \frct{1}{N(N+1)}
\:\: , \\
\label{eq:nuDef}
  \nu &\!\equiv\!& \frct{1}{N-1} - \frct{1}{N+2}
      \;\equiv\; \DNm{} - \DNpp{}
      \;=\; \frct{3}{(N-1)(N+2)}  
\; .
\eea
 
With this notation (suppressing the argument $N$ of the harmonic sums for
brevity), we can summarize the known and new all-$N$ results for 
$\gamma^{\,(3)}_{\,\rm ps}$.
The leading large-$\nf$ contribution has been published in eq.~(3.10) of 
ref.~\cite{Davies:2016jie},
\bea
\label{eq:gpsnf3}
\lefteqn{\gamma_{\,\rm ps}^{\,(3)}(N)\big|_{\colourcolour{\nft}} \;=\,
%%START
%%L texgqq3nf3 = 
          16/9\, \* \colourcolour{\cf} \, \*  \Big\{\,
            2/3\, \* \Big(2\* \Sss(1,1,1) - 3\*\zeta_3 \Big)\,\* \Big(
                9\, \* \eta\,
              + 6\, \* \eta^{2}\,
              - 4\, \* \nu
              \Big)\,
          + 4/3\, \* \Ss(1,1)\, \* \Big(11\, \* D_{0}\,
              - 13\, \* D_{0}^{2}\,
%%STOP
}
%%START
   \nn \\[0mm] & & \mbox{} \vphantom{\Big(}
              + 6\, \* D_{0}^{3}\,
              - 17\, \* D_{1}\,
              - 4\, \* D_{1}^{2}\,
              + 12\, \* D_{1}^{3}\,
              + 2\, \* D_{2}\,
              + 8\, \* D_{2}^{2}\,
              + 4\, \* D_{-1}\Big)\,
          - 2/9\, \* \S(1)\, \* \Big(
                94\, \* D_{0}\,
              - 98\, \* D_{0}^{2}\,
              + 87\, \* D_{0}^{3}\,
   \nn \\[0mm] & & \mbox{} \vphantom{\Big(}
              - 18\, \* D_{0}^{4}\,
              - 226\, \* D_{1}\,
              + 100\, \* D_{1}^{2}\,
              + 111\, \* D_{1}^{3}\,
              - 90\, \* D_{1}^{4}\,
              + 128\, \* D_{2}\,
              + 88\, \* D_{2}^{2}\,
              - 48\, \* D_{2}^{3}\,
              + 4\, \* D_{-1}\Big)\,
   \nn \\[0mm] & & \mbox{} \vphantom{\Big(}
          + 1/9\, \* \Big(
                52\, \* D_{0}\,
              - 118\, \* D_{0}^{2}\,
              + 146\, \* D_{0}^{3}\,
              - 87\, \* D_{0}^{4}\,
              + 18\, \* D_{0}^{5}\,
              - 412\, \* D_{1}\,
              + 430\, \* D_{1}^{2}
              - 54\, \* D_{1}^{3}\,
              - 309\, \* D_{1}^{4}\,
   \nn \\[0mm] & & \mbox{} \vphantom{\Big(}
              + 198\, \* D_{1}^{5}\,
              + 364\, \* D_{2}\,
              + 72\, \* D_{2}^{2}\,
              - 176\, \* D_{2}^{3}\,
              + 96\, \* D_{2}^{4}\,
              - 4\, \* D_{-1}\,
              \Big)
          \Big\}
%%;
%%STOP
\, .
\eea
Additional all-$N$ information can be obtained for the terms involving 
Riemann-$\zeta$ values, due to the reduced complexity of the harmonic sums. 
The reconstruction of the all-$N$ expressions follows the same approach as in 
the nonsinglet case in ref.~\cite{Moch:2017uml}. 
The $\zeta_3$ dependence of the quartic colour factor $\dfRRnR$ is given by 
(see app.~\ref{sec:appA} for the definition of all colour factors),
\bea
\label{eq:psdRR4ncz3}
\lefteqn{
  \left. \gamma_{\,\rm ps}^{\,(3)}(N)
  \right|_{\,\zr3\,\colourcolour{\nf\,\dfRRnR}}  \;=\,
%%START
%%L %%texgps3z3d4RRnc =
256\*\Big\{\,
  \Big(
  \S(3) + 4\* \Ss(-2,1) - 2\* \S(-3)
  \Big)\,\* \Big( 
  3 - 6\*\eta 
  \Big)\,
%%STOP
}
%%START
   \nn \\[0mm] & & \mbox{} \vphantom{\Big(}
  +
  \Big(
  2\* \Ss(1,1) - \S(2)
  \Big)\,\* \Big( 
  27\*\eta 
   + 22\*\eta^2 - 32/3\*\nu
  \Big)\,
  + \S(-2) \* \Big(  
  - {1/3} + {92/3}\*\eta + 20\*\eta^2 - 32/3\*\nu 
  \Big)\,
   \nn \\[0mm] & & \mbox{} \vphantom{\Big(}
  + \S(1) \* \Big(  
  - 70\*\eta - 68\*\eta^2 - 24\*\eta^3 + 64/3\*\nu
  \Big)\,
  + {8/3}\*\eta - 3\*\eta^2
 \Big\}
%%;
%%STOP
\; .
\eea
This is a new result based on computing the relevant OMEs for $N \leq 22$. 
Likewise, for the colour factor $\nfs\, \cfs$ the term proportional to 
$\zeta_3$ reads
\bea
\label{eq:psQEDz3}
\lefteqn{
  \left. \gamma_{\,\rm ps}^{\,(3)}(N)
  \right|_{\,\zr3\,\colourcolour{\nfs\:\!\cfs}}   \;=\,
%%START
%%L %%texgps3z3QED =
32/9\*\Big\{\,
         2\, \*\S(1)\* \,(\, 
            9\*\,\eta 
          + 6\*\,\eta^2 
          - 4\*\,\nu \,) 
       - 335\, \* D_{0}\,
       + 145\, \* D_{0}^{2}\,
       - 66\, \* D_{0}^{3}\,
%%STOP
}
%%START
   \nn \\[0mm] & & \mbox{\hspp} \vphantom{\Big(}
       + 287\, \* D_{1}\,
       + 187\, \* D_{1}^{2}\,
       + 90\, \* D_{1}^{3}\,
       + 36\, \* D_{2}\,
       + 8\, \* D_{2}^{2}\,
       + 12\, \* D_{-1}\,
       - 16\, \* D_{-1}^{2}\,
 \Big\}
%%;
%%STOP
\, ,
\eea
which is new as well. This expression has been derived from and verified
using moments up to $N = 52$, which are part of the computations of the
$\nfs$ QED contributions to $\gamma^{\,(3)}_{\,\rm ij}$ to very high $N$
\cite{MVV-unpubl}.

Next, the $\zeta_4$ part in $\gamma^{\,(3)}_{\,\rm ps}(N)$ can 
be derived with the help of the no-$\pi^2$ 
conjecture$/$theorem~\cite{Jamin:2017mul,Baikov:2018wgs}.
This has been done in ref.~\cite{Davies:2017hyl},
eq.~(9).\footnote{
The coefficients proportional to $\nu$ in eqs.~(9) and~(12) in 
ref.~\cite{Davies:2017hyl} need to be replaced as follows: $\nu \to \nu/3$.
Note further that the definition of $\nu$ in ref.~\cite{Davies:2016jie} 
differs from that in ref.~\cite{Moch:2018wjh} and the present paper by this 
factor of 3.}
Eqs.~(\ref{eq:GpsN2}) -- (\ref{eq:GpsN20}) agree with this result,
\bea
  \label{gps3z4}
\lefteqn{  \left. \gamma_{\,\rm ps}^{\,(3)}(N)
  \right|_{\,\zr4} \;=\,
%%START
%%L texgps3z4 =
16\*\,\colourcolour{\cf\*\,\camcf} \*\, \Big\{\,
         \colourcolour{\cf\*\,\nf} \*\, \Big(
            46\*\,\nu
          - 8\*\,\nu^2
          - 117\*\,\eta
          - 87\*\,\eta^2
          - 18\*\,\eta^3
          \Big)
%%STOP
}
%%START
   \nn \\[0mm] & & \mbox{} \vphantom{\Big(}
       +\, \colourcolour{\camcf\*\,\nf}\*\, \Big(
            38\*\,\nu
          - 8\*\,\nu^2
          - {195/2}\*\,\eta
          - 69\*\,\eta^2
          - 12\*\,\eta^3
          + \S(1)\, \*(\, 8\*\,\nu
          - 18\*\,\eta 
          - 12\*\,\eta^2 \,)
          \Big)
   \nn \\[0mm] & & \mbox{} \vphantom{\Big(}
       + \colourcolour{\nfs}\*\, \Big(
          15\*\,\eta
          + 10\*\,\eta^2
          - {20/3}\*\,\nu
          \Big)
\Big\}
%%;
%%STOP
\; .
\eea

Finally, with the moments up to $N = 20$ the terms with $\zeta_5$ can be
readily determined and verified for all colour factors, thus extending the 
result in eq.~(3.13) ref.~\cite{Moch:2018wjh}, where the all-$N$ $\zeta_5$ 
contributions to the quartic colour factors have been given,
\bea
\label{gps3z5}
\lefteqn{  \left. \gamma_{\,\rm ps}^{\,(3)}(N)
  \right|_{\,\zr5} \;=\,
%%START
%%L %%texgps3z5 =
160\,\*\colourcolour{\nf\, \*\cft\,}\*\, \Big(
       9\, \* \eta
     + 6\, \* \eta^2
     - 4\, \* \nu
\Big)
+ {80/3}\,\* \colourcolour{\nf\, \*\ca\, \*\cfs\,}\*\, \Big(
     - 9\, \* \eta
     - 6\, \* \eta^2
     + 4\, \* \nu
\Big)
%%STOP
}
%%START
   \nn \\[0mm] & & \mbox{\hspp} \vphantom{\Big(}
+ {40/9}\,\* \colourcolour{\nf\, \*\cas\, \*\cf\,}\*\, \Big(
     - 1
     - 214\, \* \eta
     - 144\, \* \eta^2
     + 104\, \* \nu
\Big)
   \nn \\[0mm] & & \mbox{\hspp} \vphantom{\Big(}
+ {320/3}\,\* \colourcolour{\nf\, \*\dfRRnr}\*\, \Big(
     - 1
     + 56\, \* \eta
     + 36\, \* \eta^2
     - 16\, \* \nu
\Big)
%%;
%%STOP
\: .
\hspace*{40mm}\,
\eea
Expressions in $x$-space for the leading large-$\nf$ part of
eq.~(\ref{eq:gpsnf3}) have been presented in ref.~\cite{Davies:2016jie}, 
eq.~(4.21).
On the other hand, the $N$-space terms with Riemann-$\zeta$ values do not
correspond to the $x$-space contributions with Riemann-$\zeta$ values, 
as the inverse Mellin transformation generates additional terms with $\zeta_n$.
Similarly, it is not possible to read off the coefficients of $\zeta_n$
in the limit $N \to \infty\,$ from eqs.~(\ref{eq:psdRR4ncz3}) -- 
(\ref{gps3z5}), as non-$\zeta$ harmonic sums contribute to these.

\begin{figure}[p]
\vspace{-4mm}
\centerline{\epsfig{file=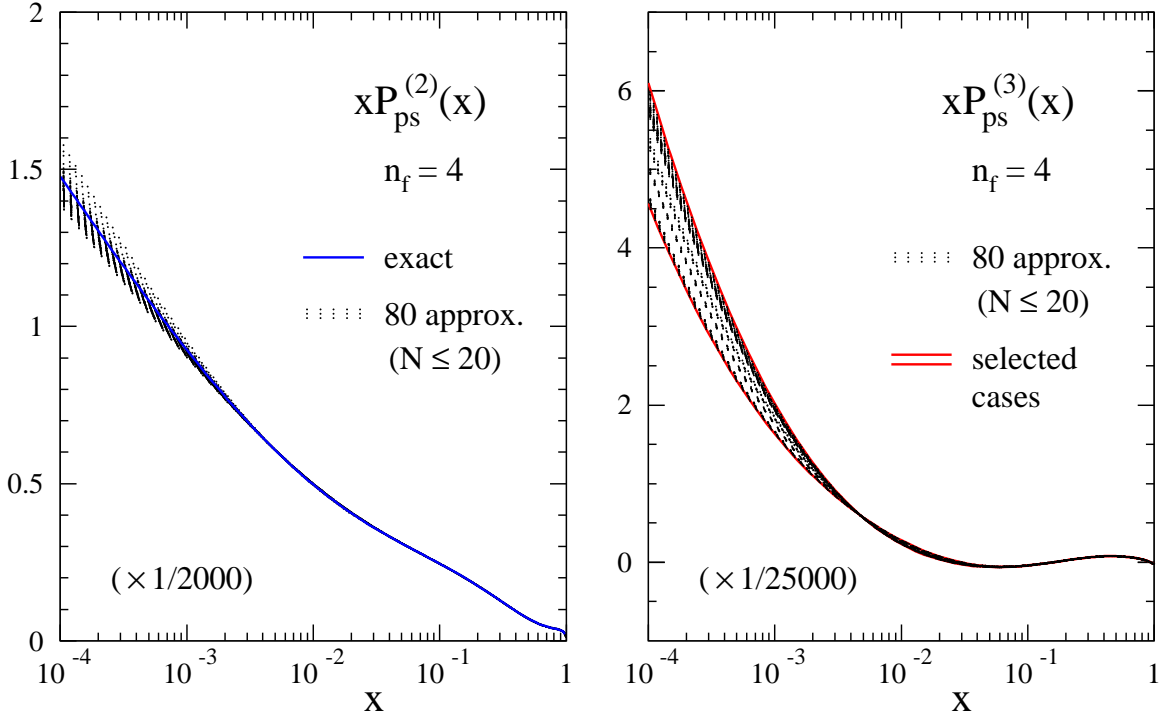,width=16.0cm,angle=0}}
\vspace{-2mm}
\caption{ \label{fig:P3psa} \small
 80 trial functions, constructed as described in the text, for the splitting 
 functions $P_{\rm ps}^{\,(n)}(x)$ at $\nf=4$.
 At $n=2$ (left panel) the known exact result is shown by the solid (blue) 
 line.  At $n=3$ (right panel) two functions, shown by the solid (red) lines, 
 are chosen to represent the remaining uncertainty.
 }
\vspace{1mm}
\end{figure}

For phenomenology applications, the moments in eq.~(\ref{eq:gps3-numerics})
can be used to construct approximate representations for the $\nfo$ and $\nfs$ 
parts of $P_{\,\rm ps}^{\,(3)}(x)$, subject to the constraints imposed 
by the known terms in the limits $x \ra 0,\:1$.
At small $x$, the coefficient of the leading logarithm $(\ln^2x)/x$ is 
known since long~\cite{Catani:1994sq}, as well as those of the highest three 
sub-dominant logarithms $\ln^{\,k}x$ with $k=6,5,4$, see 
ref.~\cite{Davies:2022ofz}.
At large $x$, the leading terms are of the form 
$(1-x)^j\ln^{\,k}(1-x)$ with $j \geq 1$ and $k \leq 4$. 
The coefficients for $k=4,3$ are known~\cite{Soar:2009yh} for all $j$.
With the 10 Mellin moments $N \leq 20$ in eq.~(\ref{eq:gps3-numerics}), 
the coefficients of all remaining unknown small-$x$ and large-$x$ 
(for $j\!=\!1$) terms can be `fitted' together with a three-parameter 
interpolating function.
Thus all approximations include
\begin{itemize}
\item the next-to-leading and next-to-next-to-leading
 small-$x$ terms: $(\ln x)/x$ and $1/x$,\\[-7mm]
\item the remaining three sub-dominant small-$x$ logarithms: 
 $\ln^{\,k}x$ with $k=3,2,1$,\\[-7mm]
\item the two remaining $j=1$ large-$x$ terms: $(1-x)\ln^k(1-x)$ with $k=2,1$.
\end{itemize}
Choosing 10 two-parameter polynomials together one function that includes
$\ln^{\,k}(1-x)$ with $k=1$ or $2$ or the dilogarithm Li$_2(x)$ (suitably 
suppressed as $x \ra 1$), we have thus build 80 trial functions that fulfil 
all known constraints. 
These functions are shown $\nf=4$ by the black-dotted curves in the right 
part of fig.~\ref{fig:P3psa}; other values of $\nf$ show qualitatively the 
same behaviour.
As a check, exactly the same procedure has been applied to the NNLO splitting 
function $P_{\rm ps}^{\,(2)}(x)$, where it can be compared with the exact 
result \cite{Vogt:2004mw}; this comparison is shown in the left part of the 
figure. 

\begin{figure}[p]
\vspace{-2mm}
\centerline{\epsfig{file=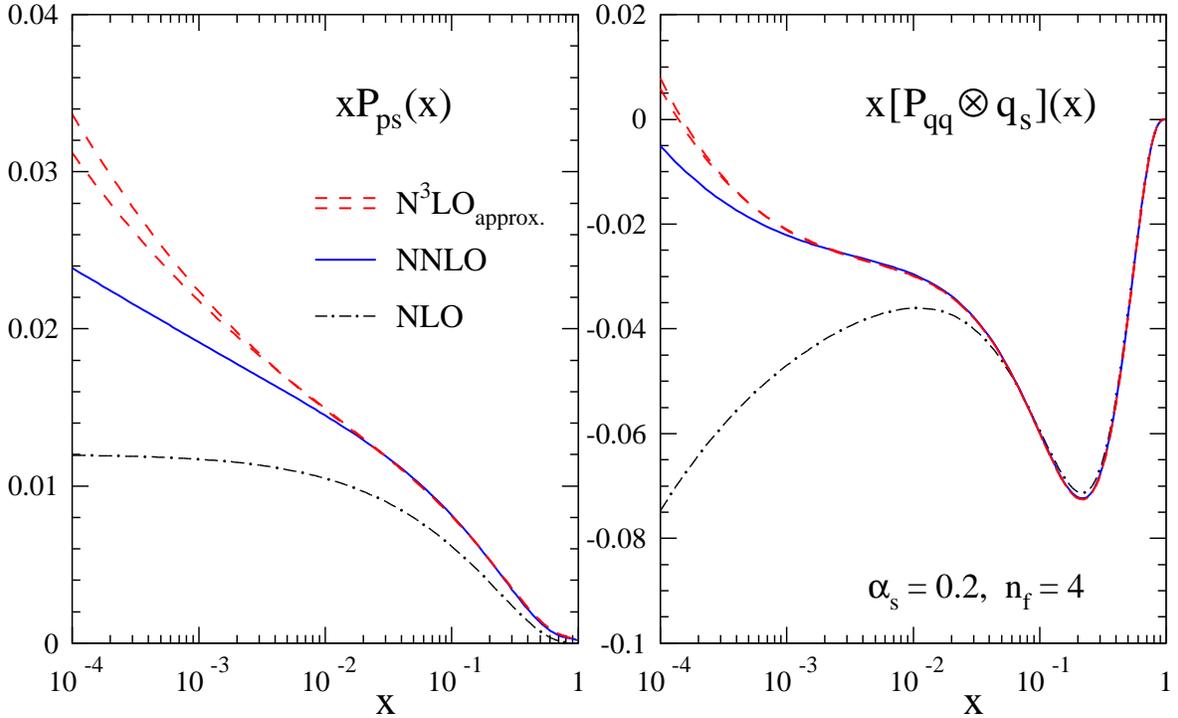,width=16.0cm,angle=0}}
\vspace{-2mm}
\caption{\small \label{fig:Pxqs}
 Left: The NLO, NNLO and N$^3$LO approximations for $P_{\,\rm ps}(x)$
 for a fixed value $\als (\mu_{0}^{\,2}) \; = \; 0.2$ of the strong coupling
 and $\nf =4$.
 Right: The resulting perturbative expansion of the contribution of 
 $P_{\,\rm qq}$ to the scale derivative of the singlet quark PDF 
 $q_{\rm s}^{}$ in eq.~(\ref{eq:sgEvol}) for the initial distribution 
 (\ref{eq:qs-shape}).
 }
\end{figure}

Since our approximation procedure passes this test (and others), we have
selected two representatives for each physically relevant value of $\nf$
that indicate the remaining uncertainty in $P_{\rm ps}^{\,(3)}(x)$.
The chosen approximation, shown in red in fig.~\ref{fig:P3psa}, are,
with $x_1=1\!-\!x$, $L_1=\ln (1\!-\!x)$ and $L_0=\ln x$,
\bea
\label{eq:Pps3A3-nf3}
{\lefteqn{ 
 P_{\rm ps,\,A}^{\,(3)}(\nf=3,x) \; = \; 
 p_{{\rm ps},0}^{\,(\nf=3)}(x) + 
 67731\, \*x_1\*L_0/x
+274100\, \*x_1/x
+40006\, \*L_0^3
+10620\, \*L_0^2
}}
\nn \\ && \mbox{}
+353656\, \*x_1\*L_0
-2365.1\, \*x_1\*L_1^2
-7412.1\, \*x_1\*L_1
+1533.0\, \*x_1^2\*L_1^2
-104493\, \*x_1\*(1\!+\!2\*x)
+34403\, \*x_1\*x^2
\, ,
\nn\\
{\lefteqn{ 
 P_{\rm ps,\,B}^{\,(3)}(\nf=3,x) \; = \; 
 p_{{\rm ps},0}^{\,(\nf=3)}(x) + 
 54593\, \*x_1\*L_0/x
+179748\, \*x_1/x
-2758.3\, \*L_0^3
-103604\, \*L_0^2
}}
\nn \\ && \mbox{}
+4700.0\, \*x_1\*L_0
-1986.9\, \*x_1\*L_1^2
-2801.2\, \*x_1\*L_1
-6005.9\, \*x_1^2\*L_1^2
-195263\, \*x_1
+12789\, \*x_1\*x\*(1+x)
\, ,
\nn\\
\\
\label{eq:Pps3A3-nf4}
{\lefteqn{ 
 P_{\rm ps,\,A}^{\,(3)}(\nf=4,x) \; = \; 
 p_{{\rm ps},0}^{\,(\nf=4)}(x) + 
 90154\,\*x_1\*L_0/x
+359084\,\*x_1/x
+52525\,\*L_0^3
+13869\,\*L_0^2
}}
\nn \\ && \mbox{}
+461167\,\*x_1\*L_0
-2491.5\,\*x_1\*L_1^2
-7498.2\,\*x_1\*L_1
+1727.2\,\*x_1^2\*L_1^2
-136319\,\*x_1\*(1\!+\!2\*x)
+45379\,\*x_1\*x^2
\, ,
\nn\\
{\lefteqn{ 
 P_{\rm ps,\,B}^{\,(3)}(\nf=4,x) \; = \; 
 p_{{\rm ps},0}^{\,(\nf=4)}(x) + 
 72987\,\*x_1\*L_0/x
+235802\,\*x_1/x
-3350.9\,\*L_0^3
-135378\,\*L_0^2
}}
\nn \\ && \mbox{}
+5212.9\,\*x_1\*L_0
-1997.2\,\*x_1\*L_1^2
-1472.7\,\*x_1\*L_1
-8123.3\,\*x_1^2\*L_1^2
-254921\,\*x_1
+17138\,\*x_1\*x\*(1+x)
\, ,
\nn\\
\\
\label{eq:Pps3A3-nf5}
{\lefteqn{ 
 P_{\rm ps,\,A}^{\,(3)}(\nf=5,x) \; = \; 
 p_{{\rm ps},0}^{\,(\nf=5)}(x) + 
 112481\,\*x_1\*L_0/x
+440555\,\*x_1/x
+64577\,\*L_0^3
+16882\,\*L_0^2
}}
\nn \\ && \mbox{}
+562992\,\*x_1\*L_0
-2365.7\,\*x_1\*L_1^2
-6570.1\,\*x_1\*L_1
+1761.7\,\*x_1^2\*L_1^2
-166581\,\*x_1\*(1\!+\!2\*x)
+56087\,\*x_1\*x^2
\, ,
\nn\\
{\lefteqn{ 
 P_{\rm ps,\,B}^{\,(3)}(\nf=5,x) \; = \; 
 p_{{\rm ps},0}^{\,(\nf=5)}(x) + 
 91468\,\*x_1\*L_0/x
 +289658\,\*x_1/x
 -3814.9\,\*L_0^3
 -165795\,\*L_0^2
}}
\nn \\ && \mbox{}
 +4908.9\,\*x_1\*L_0
 -1760.8\,\*x_1\*L_1^2
 +804.5\,\*x_1\*L_1
 -10295\,\*x_1^2\*L_1^2
 -311749\,\*x_1
 +21521\,\*x_1\*x\*(1+x)
\, ,
\eea
with the known endpoint contribution, with coefficients rounded to seven 
significant figures,
\bea
\label{eq:Pps30-nf}
{\lefteqn{ \hspn
 p_{{\rm ps},0}^{\,(\nf)}(x) \; = \; \nf\*\Big\{
    1749.227\,\*L_0^2/x
 -(7.506173
  -0.7901235\,\*\nf)\*L_0^6 
 }}
\nn \\ && \mbox{}
  +(28.54979
   +3.792593\,\*\nf)\*L_0^5
 -(854.8001
   -77.36626\,\*\nf
  +0.1975309\,\*\nfs)\*L_0^4
\nn \\[0.5mm] && \mbox{}
 -(199.1111
   -13.69547\,\*\nf)\*x_1^2\*L_1^3
   -13.16872\,\*x_1^2\*L_1^4
 -(247.5505
   -40.55967\,\*\nf
\nn \\ && \mbox{} 
   +1.580247\,\*\nfs)\*x_1\*L_1^3
 -(56.46091
   -3.621399\,\*\nf)\*x_1\*L_1^4
\Big\}
\, .
\eea
Specifically, we note that eqs.~(\ref{eq:Pps3A3-nf3}) -- (\ref{eq:Pps3A3-nf5})
include a numerical prediction, with a precision of $\pm 10\%$, of the
so far uncalculated coefficient of the next-to-leading small-$x$ logarithm 
$(\ln x)/x$ of $P_{\,\rm ps}^{\,(3)}(x)$ for $\nf=3,4,5$.
The above uncertainty bands also lead to the following predictions for the
$\gamma_{\,\rm ps}^{\,(3)}(N)$ at $N=22$ 
(with brackets indicating the error on the last digits), 
\bea
\label{eq:gps3N22approx}
  \gamma_{\,\rm ps}^{\,(3)}(N\!=\!22) \,=\, 
   6.2478570(6)\, , \;\;
  10.5202730(8)\, , \;\;
  15.6913948(10)
\quad \mbox{for} \quad \nf\,=\,3,4,5
\: .
\eea
 
The numerical implications for the evolution of $q_{\rm s}^{}$ due to the 
quark-quark splitting functions in eq.~(\ref{eq:Pps3A3-nf4}) are illustrated 
in fig.~\ref{fig:Pxqs} for our default value $\nf =4$.
We show the perturbative expansion for $P_{\,\rm ps}$ alone and for the 
convolution $P_{\,\rm qq} \otimes q_{\,\rm s}$ through N$^3$LO, in the latter
case using the same schematic (order-independent) model input 
\cite{Vogt:2004mw} for the singlet quark PDF in eq.~(\ref{eq:sgPDFs}),
\beq
\label{eq:qs-shape}
  xq_{\,\rm s}(x,\mu_{0}^{\,2}) \; = \; 0.6\, x^{\, -0.3} (1-x)^{3.5} 
  \left(1 + 5.0 \,x^{\, 0.8}\right) 
  \, ,
\eeq
together with $\als (\mu_{0}^{\,2}) \; = \; 0.2$.
The latter value of the strong coupling corresponds to scales in the range 
$\mu_{0}^{\,2} \,\simeq\, 25\ldots 50$ GeV$^2$ for $\als (M_Z^{\, 2}) 
= 0.114 \ldots 0.120$ beyond the leading order. 
This fixed input facilitates a direct comparison of effects of the various
perturbative order for the splitting function 
$P_{\rm qq} = P_{\rm ns}^{\,+} + P_{\rm ps}^{}$, where the four-loop results 
for $P_{\rm ns}^{(3)+}$ have been taken from refs.~\cite{Moch:2017uml,MVV-tba}.

Fig.~\ref{fig:Pxqs} shows that the convolution of the splitting functions 
with $q_{\rm s}^{}$ dampens the residual \mbox{small-$x$} uncertainties 
induced by the approximations $A$ and $B$ in eq.~(\ref{eq:Pps3A3-nf4}).
The uncertainties in these convolutions are practically negligible down to 
$ x\lsim 10^{\,-3}$ and, even if our error band were to underestimate
the uncertainty by, say, a factor of 2, perfectly tolerable even at $x$ as 
small as $x \approx 10^{\,-4}$.
Further phenomenological studies, such as the scale stability of the parton 
evolution in eq.~(\ref{eq:sgEvol}) under the variation of the renormalization 
scales require additional information at four loops on the other splitting 
functions $P_{\rm qg}$, $P_{\rm gq}$ and $P_{\rm gg}$. 
These will be subject of forthcoming publications.

With the four-loop results for the pure-singlet quark splitting function
$P_{\,\rm ps}$ presented here, we have provided a major step towards improving 
the accuracy of the flavour-singlet parton evolution by one perturbative order 
beyond the current state of the art.
The evolution of PDFs at N$^3$LO is expected to achieve percent-level 
precision, and a brief phenomenological study of our results is consistent 
with this expectation.
We have shown that the knowledge of 10 Mellin moments of 
$P^{\,(3)}_{\,\rm ps}(x)$, together with the present knowledge of the 
behavior at the endpoints $x\to 1$ and $x\to 0$, is sufficient for the 
construction of approximations that display negligible residual uncertainties 
in a wide kinematic range of parton momentum fractions $x$ probed at current 
and future colliders.
%
% ---------------------------------------------------------------------
%
\vspace*{-1mm}
\subsection*{Acknowledgements}
\vspace*{-1mm}
This work has been supported by 
the Vidi grant 680-47-551 of the Dutch Research Council (NWO),
the UKRI FLF Mr/S03479x/1;
the Consolidated Grant {\it Particle Physics at the Higgs Centre}
of the UK Science and Technology Facilities Council (STFC);
the ERC Starting Grant 715049 {\it QCDforfuture};
the Deutsche Forschungsgemeinschaft through the Research Unit FOR 2926, 
{\it Next Generation pQCD for Hadron Structure: Preparing for the EIC}, 
project number 40824754, and DFG grant MO~1801/4-2;
and the STFC Consolidated Grant ST/T000988/1.

{\footnotesize

%\bibliographystyle{JHEP}
%\bibliography{psbib}

\providecommand{\href}[2]{#2}\begingroup\raggedright\endgroup

}

\appendix
%
% ---------------------------------------------------------------------
%
\renewcommand{\theequation}{\ref{sec:appA}.\arabic{equation}}
\setcounter{equation}{0}
\section{Mellin moments of $P^{\,(3)}_{\,\rm ps}$}
\label{sec:appA}

Here we provide the exact results for the four-loop pure-singlet anomalous 
dimensions $\gamma^{\,(3)}_{\,\rm ps}(N)$ at even $N \leq 20$ for a general 
compact simple gauge group.
The numerical values in QCD, i.e., SU($\nc=3$) have been given in 
eq.~(\ref{eq:gps3-numerics}).
The colour factors are $\ca = \nc$ and $\cf = (\ncs-1)/(2\nc)$ 
for the quadratic Casimir invariants in SU$(\nc)$.
The relevant quartic colour factor $\dfRRnR$ is obtained from the 
symmetrized trace of four generators $T_r^a$, 
see, e.g., \cite{Herzog:2017ohr,Moch:2017uml,Moch:2018wjh},
\beq
\label{eq:d4def}
  d_{r}^{\,abcd} \; =\; \frac{1}{6}\: {\rm Tr} \, ( \, 
   T_{r}^{a\,} T_{r}^{b\,} T_{r}^{c\,} T_{r}^{d\,}
   + \,\mbox{ five $bcd$ permutations}\, ) 
   \; .
\eeq
with
\beq
\label{eq:d4SUn}
  \dfRRnr  \;=\;
  \frac{1}{96\,\nct}\: ( \ncs - 1 ) ( \ncf - 6\,\ncs + 18 )
\eeq
for the fundamental representation. In QCD:
$\ca=3$, $\cf=4/3$ and $\dfRRnR = 5/36$.
{\small{
\bea
\label{eq:GpsN2}
 \lefteqn{ \hspn \gamma_{\,\rm ps}^{\,(3)}(N\!=\!2) \:\equal\:
%%START
%%L %%texGpsN2 =
    \nf\, \* \cft \, \* \left( 
      { 227938 \over 2187 }
    + { 1952 \over 81 }\, \* \zeta_3
    + { 256 \over 9 }\, \* \zeta_4
    - { 640 \over 3 }\, \* \zeta_5 \right)
} \nn \\[0.5mm] &&  \mbox{\hspn} 
  + \,\nf\, \* \ca \* \cfs \, \* \left(
    - { 162658 \over 6561 }
    + { 8048 \over 27 }\, \* \zeta_3
    - { 1664 \over 9 }\, \* \zeta_4
    + { 320 \over 9 }\, \* \zeta_5 \right)
\nn \\[0.5mm] && \mbox{\hspn}
  + \,\nf\, \* \cas\, \* \cf \, \*  \left(
    - { 410299 \over 6561 }
    - { 26896 \over 81 }\, \* \zeta_3
    + { 1408 \over 9 }\, \* \zeta_4
    + { 4480 \over 27 }\, \* \zeta_5 \right)
\nn \\[0.5mm] && \mbox{\hspn}
  + \,\nf \, \* \dfRRnr \, \* \left( 
      { 1024 \over 9 }
    + { 256 \over 9 }\, \* \zeta_3
    - { 2560 \over 9 }\, \* \zeta_5 \right)
  + \,\nfs\, \* \cfs \, \*  \left(
    - { 73772 \over 6561 }
    - { 5248 \over 81 }\, \* \zeta_3
    + { 320 \over 9 }\, \* \zeta_4 \right)
\nn \\[0.5mm] && \mbox{\hspn}
  + \,\nfs\, \* \ca \* \cf \, \* \left( 
      { 160648 \over 6561 } + 48\, \* \zeta_3
    - { 320 \over 9 }\, \* \zeta_4 \right)
  + \nft\, \* \cf \, \*  \left(
    - { 1712 \over 729 }
    + { 128 \over 27 }\, \* \zeta_3 \right)
%%;
%%STOP
\; , \\[2mm]
\label{eq:GpsN4}
 \lefteqn{ \hspn \gamma_{\,\rm ps}^{\,(3)}(N\!=\!4) \:\equal\:
%%START
%%L %%texGpsN4 =
    \nf\, \* \cft \, \* \left( 
      { 1995890620891 \over 52488000000 }
    - { 897403 \over 202500 }\, \* \zeta_3
    + { 18997 \over 2250 }\, \* \zeta_4
    - { 484 \over 15 }\, \* \zeta_5 \right)
} \nn \\[0.5mm] &&  \mbox{\hspn}
  + \,\nf\, \* \ca\, \* \cfs \, \* \left( 
      { 209865827521 \over 26244000000 }
    + { 6743539 \over 202500 }\, \* \zeta_3
    - { 29161 \over 750 }\, \* \zeta_4
    + { 242 \over 45 }\, \* \zeta_5 \right)
\nn \\[0.5mm] && \mbox{\hspn}
  + \,\nf\, \* \cas\, \* \cf \, \*  \left(
    - { 55187654921 \over 3280500000 }
    - { 3104267 \over 67500 }\, \* \zeta_3
    + { 34243 \over 1125 }\, \* \zeta_4
    + { 3164 \over 135 }\, \* \zeta_5 \right)
\nn \\[0.5mm] && \mbox{\hspn}
  + \,\nf\, \* \dfRRnr \, \* \left( 
      { 172231 \over 675 }
    - { 5368 \over 25 }\, \* \zeta_3
    - { 3728 \over 45 }\, \* \zeta_5 \right)
  + \,\nfs\, \* \cfs \, \*  \left(
    - { 141522185707 \over 26244000000 }
    - { 1207 \over 135 }\, \* \zeta_3
    + { 242 \over 45 }\, \* \zeta_4 \right)
\nn \\[0.5mm] && \mbox{\hspn}
  + \,\nfs\, \* \ca\, \* \cf \, \* \left( 
      { 9398360351 \over 1640250000 }
    + { 57877 \over 10125 }\, \* \zeta_3
    - { 242 \over 45 }\, \* \zeta_4 \right)
  + \,\nft\, \* \cf \, \*  \left(
    - { 46099151 \over 72900000 }
    + { 484 \over 675 }\, \* \zeta_3 \right)
%%;
%%STOP
\; , \\[2mm]
\label{eq:GpsN6}
 \lefteqn{ \hspn \gamma_{\,\rm ps}^{\,(3)}(N\!=\!6) \:\equal\:
%%START
%%L %%texGpsN6 =
    \nf\, \* \cft \, \* \left( 
      { 140565274663259489 \over 5403265623000000 }
    - { 62727544 \over 24310125 }\, \* \zeta_3
    + { 343156 \over 77175 }\, \* \zeta_4
    - { 1936 \over 147 }\, \* \zeta_5 \right)
} \nn \\[0.5mm] &&  \mbox{\hspn}
  + \,\nf\, \* \ca\, \* \cfs \, \* \left( 
      { 336481838777617 \over 360217708200000 }
    + { 2111992 \over 324135 }\, \* \zeta_3
    - { 1389806 \over 77175 }\, \* \zeta_4
    + { 968 \over 441 }\, \* \zeta_5 \right)
\nn \\[0.5mm] && \mbox{\hspn}
  + \,\nf\, \* \cas\, \* \cf \, \*  \left(
    - { 6194882229735067 \over 864522499680000 }
    - { 2396237 \over 165375 }\, \* \zeta_3
    + { 41866 \over 3087 }\, \* \zeta_4
    + { 9544 \over 1323 }\, \* \zeta_5 \right)
\nn \\[0.5mm] && \mbox{\hspn}
  + \,\nf\, \* \dfRRnr \, \* \left( 
      { 64697569 \over 330750 }
    - { 426976 \over 3675 }\, \* \zeta_3
    - { 39808 \over 441 }\, \* \zeta_5 \right)
\nn \\[0.5mm] && \mbox{\hspn}
  + \,\nfs\, \* \cfs \, \*  \left(
    - { 812984663253277 \over 270163281150000 }
    - { 2594876 \over 694575 }\, \* \zeta_3
    + { 968 \over 441 }\, \* \zeta_4 \right)
\nn \\[0.5mm] && \mbox{\hspn}
  + \,\nfs\, \* \ca\, \* \cf \, \* \left( 
      { 3092531515013 \over 964868861250 }
    + { 217432 \over 99225 }\, \* \zeta_3
    - { 968 \over 441 }\, \* \zeta_4 \right)
  + \,\nft\, \* \cf \, \*  \left(
    - { 19597073837 \over 61261515000 }
    + { 1936 \over 6615 }\, \* \zeta_3 \right)
%%;
%%STOP
\; , \\[2mm]
\label{eq:GpsN8}
 \lefteqn{ \hspn \gamma_{\,\rm ps}^{\,(3)}(N\!=\!8) \:\equal\:
%%START
%%L %%texGpsN8 =
%
       \nf\, \*\cft\,  \* \left(
            {3960340604223955458923 \over 192072198786048000000}
          - {34718701049 \over 18003384000}\*\zeta_3
          + {13529827 \over 4762800}\*\zeta_4
          - {1369 \over 189}\*\zeta_5
          \right)
} \nn \\[0.5mm] &&  \mbox{\hspn}
       + \,\nf\, \*\ca\, \*\cfs\,  \* \left(
          - {43838488788848637899 \over 13719442770432000000}
          + {10167760657 \over 18003384000}\*\zeta_3
          - {10211371 \over 952560}\*\zeta_4
          + {1369 \over 1134}\*\zeta_5
          \right)
\nn \\[0.5mm] && \mbox{\hspn}
       + \,\nf\, \*\cas\, \*\cf\,  \* \left(
          - {8552512702477166383 \over 2939880593664000000}
          - {97528710971 \over 18003384000}\*\zeta_3
          + {1340251 \over 170100}\*\zeta_4
          + {128 \over 63}\*\zeta_5
          \right)
\nn \\[0.5mm] && \mbox{\hspn}
       + \,\nf\, \*\dfRRnr\,  \* \left(
            {1183211180737 \over 7715736000}
          - {18321694 \over 297675}\*\zeta_3
          - {18164 \over 189}\*\zeta_5
          \right)
\nn \\[0.5mm] && \mbox{\hspn}
       + \,\nfs\, \*\cfs\, \* \left(
          - {5115927245667479753 \over 2743888554086400000}
          - {15129691 \over 7144200}\*\zeta_3
          + {1369 \over 1134}\*\zeta_4
          \right)
\nn \\[0.5mm] && \mbox{\hspn}
       + \,\nfs\, \*\ca\, \*\cf\, \* \left(
            {15301312238130101 \over 7349701484160000}
          + {8397097 \over 7144200}\*\zeta_3
          - {1369 \over 1134}\*\zeta_4
          \right)
\nn \\[0.5mm] && \mbox{\hspn}
       + \,\nft\, \*\cf\, \* \left(
          - {162840799744061 \over 816633498240000}
          + {1369 \over 8505}\*\zeta_3
          \right)
%
%%;
%%STOP
\; , \\[2mm]
\label{eq:GpsN10}
 \lefteqn{ \hspn \gamma_{\,\rm ps}^{\,(3)}(N\!=\!10) \:\equal\:
%%START
%%L %%texGpsN10 =
%
         \nf\, \* \cft\, \* \left(
            {19206657411733877390649313 \over 1118944450162341495000000}
          - {45224548192 \over 28017383625}\*\zeta_3
          + {1080128 \over 539055}\*\zeta_4
          - {25088 \over 5445}\*\zeta_5
          \right)
} \nn \\[0.5mm] &&  \mbox{\hspn}
       + \,\nf\, \*\ca\, \* \cfs\, \* \left(
          - {1538138456874500390560463 \over 298385186709957732000000}
          - {31074715888 \over 28017383625}\*\zeta_3
          - {97295744 \over 13476375}\*\zeta_4
          + {12544 \over 16335}\*\zeta_5
          \right)
\nn \\[0.5mm] && \mbox{\hspn}
       + \,\nf\, \*\cas\, \*\cf\, \* \left(
          - {202179113304531644762417 \over 284176368295197840000000}
          - {192321673117627 \over 109828143810000}\*\zeta_3
          + {23430848 \over 4492125}\*\zeta_4
          - {14912 \over 49005}\*\zeta_5
          \right)
\nn \\[0.5mm] && \mbox{\hspn}
       + \,\nf\, \* \dfRRnr \,\* \left(
            {1240606813603 \over 9901861200}
          - {182828576543 \over 6303268125}\*\zeta_3
          - {1624576 \over 16335}\*\zeta_5
          \right)
\nn \\[0.5mm] && \mbox{\hspn}
       + \,\nfs\, \* \cfs\,\*\left(
          - {367710354086746558213 \over 296017050307497750000}
          + {12544 \over 16335}\*\zeta_4
          - {1243744 \over 898425}\*\zeta_3
          \right)
\nn \\[0.5mm] && \mbox{\hspn}
       + \,\nfs\,\*\ca\,\*\cf\, \* \left(
            {314242565140920849001 \over 215285127496362000000}
          - {12544 \over 16335}\*\zeta_4
          + {89550464 \over 121287375}\*\zeta_3
          \right)
\nn \\[0.5mm] && \mbox{\hspn}
       + \,\nft\, \* \cf\, \* \left(
          - {2205751150439 \over 15885856515375}
          + {25088 \over 245025}\*\zeta_3
          \right)
%
%%;
%%STOP
\; , \\[2mm]
\label{eq:GpsN12}
 \lefteqn{ \hspn \gamma_{\,\rm ps}^{\,(3)}(N\!=\!12) \:\equal\:
%%START
%%L %%texGpsN12 =
%
       \nf\, \*\cft\, \* \left(
            {88961716829219432715740321165467 \over 6065706909620362869213148800000}
          - {671898176890091 \over 475272972023850}\*\zeta_3
\right.} \nn \\[0.5mm] &&  \mbox{\hspp}\left.
          + {2642589184 \over 1758511755}\*\zeta_4
          - {124820 \over 39039}\*\zeta_5
          \right)
\nn \\[0.5mm] &&  \mbox{\hspn}
       + \,\nf\, \*\ca\, \* \cfs\, \* \left(
          - {116621076523257514706541796876157 \over 19410262110785161181482076160000}
          - {1989733300788683 \over 1267394592063600}\*\zeta_3
\right.\nn \\[0.5mm] &&  \mbox{\hspp}\left.
          - {18419693641 \over 3517023510}\*\zeta_4
          + {62410 \over 117117}\*\zeta_5
          \right)
\nn \\[0.5mm] && \mbox{\hspn}
       + \,\nf\, \*\cas\, \*\cf\, \* \left(
            {108542242435054542124290045599 \over 210067771761744168630758400000}
          + {7458806358343849 \over 228131026571448000}\*\zeta_3
\right.\nn \\[0.5mm] &&  \mbox{\hspp}\left.
          + {108549713 \over 29066310}\*\zeta_4
          - {549260 \over 351351}\*\zeta_5
          \right)
\nn \\[0.5mm] && \mbox{\hspn}
       + \,\nf\, \* \dfRRnr \* \left(
            {13789024918875535939 \over 130167745507800000}
          - {261789233833 \over 33202669500}\*\zeta_3
          - {11883280 \over 117117}\*\zeta_5
          \right)
\nn \\[0.5mm] && \mbox{\hspn}
       + \,\nfs\, \*\cfs\, \* \left(
          - {2797424774494087428631891051 \over 3209368735248869242969920000}
          - {15566147588 \over 15826605795}\*\zeta_3
          + {62410 \over 117117}\*\zeta_4
          \right)
\nn \\[0.5mm] && \mbox{\hspn}
       + \,\nfs\, \* \ca\, \* \cf\, \* \left(
            {17400519563132679535658867 \over 16159059366288012971596800}
          + {536686847 \over 1055107053}\*\zeta_3
          - {62410 \over 117117}\*\zeta_4
          \right)
\nn \\[0.5mm] && \mbox{\hspn}
       + \,\nft\, \*\cf\, \* \left(
          - {127821768039445576087 \over 1233139451029305019200}
          + {24964 \over 351351}\*\zeta_3
          \right)
%
%%;
%%STOP
\; , \\[2mm]
\label{eq:GpsN14}
 \lefteqn{ \hspn \gamma_{\,\rm ps}^{\,(3)}(N\!=\!14) \:\equal\:
%%START
%%L %%texGpsN14 =
%
       \nf\, \*\cft\, \* \left(
            {378715964141637885273854172708551 \over 29692271585554223835309120000000}
          - {367639947406454 \over 290813881483125}\*\zeta_3
\right.} \nn \\[0.5mm] &&  \mbox{\hspp}\left.
          + {505959889 \over 430404975}\*\zeta_4
          - {22472 \over 9555}\*\zeta_5
          \right)
\nn \\[0.5mm] && \mbox{\hspn}
       + \,\nf\, \* \ca\, \*\cfs\, \* \left(
          - {158391038218926832370900571697207 \over 25124229803161266322184640000000}
          - {14044584522181 \over 8616707599500}\*\zeta_3
\right.\nn \\[0.5mm] &&  \mbox{\hspp}\left.
          - {17190217637 \over 4304049750}\*\zeta_4
          + {11236 \over 28665}\*\zeta_5
          \right)
\nn \\[0.5mm] && \mbox{\hspn}
       + \,\nf \* \cas\, \* \cf\, \* \left(
          + {212618793832045564739311832977 \over 171789605491700966305536000000}
          + {77579129461513987 \over 76774864711545000}\*\zeta_3
\right.\nn \\[0.5mm] &&  \mbox{\hspp}\left.
          + {933124519 \over 331080750}\*\zeta_4
          - {15356 \over 6615}\*\zeta_5
          \right)
\nn \\[0.5mm] && \mbox{\hspn}
       + \,\nf\, \*\dfRRnr\, \* \left(
          + {19073114986773056430079 \over 207292134721171500000}
          + {1655601644872528 \over 246073284331875}\*\zeta_3
          - {2945344 \over 28665}\*\zeta_5
          \right)
\nn \\[0.5mm] && \mbox{\hspn}
       + \,\nfs\, \*\cfs\, \* \left(
          - {4032068581057610850590942023 \over 6344502475545774323784000000}
          - {14300095639 \over 19368223875}\*\zeta_3
          + {11236 \over 28665}\*\zeta_4
          \right)
\nn \\[0.5mm] && \mbox{\hspn}
       + \,\nfs\, \* \ca\, \* \cf\, \* \left(
          + {15677247599879616342540623 \over 19014492933703619352000000}
          + {72862649 \over 195638625}\*\zeta_3
          - {11236 \over 28665}\*\zeta_4
          \right)
\nn \\[0.5mm] && \mbox{\hspn}
       + \,\nft\, \*\cf\, \* \left(
          - {1225131890207918292167 \over 15090867407701285200000}
          + {22472 \over 429975}\*\zeta_3
          \right)
%
%%;
%%STOP
\; , \\[2mm]
\label{eq:GpsN16}
 \lefteqn{ \hspn \gamma_{\,\rm ps}^{\,(3)}(N\!=\!16) \:\equal\:
%%START
%%L %%texGpsN16 =
%
        \nf\, \* \cft\, \* \left(
            {52845922593469053066814397892836049514811 \over 4700647014054801534593682386898124800000}
          - {1576051471675106357 \over 1378077594461568000}\*\zeta_3
\right.} \nn \\[0.5mm] &&  \mbox{\hspp}\left.
          + {1209091491169 \over 1274723049600}\*\zeta_4
          - {18769 \over 10404}\*\zeta_5
          \right)
\nn \\[0.5mm] && \mbox{\hspn}
       + \,\nf\, \*\ca\, \*\cfs\, \* \left(
          - {272131316618720437180274758003365964006181 \over 43089264295502347400442088546566144000000}
          - {186362119838618569 \over 120140097978700800}\*\zeta_3
\right.\nn \\[0.5mm] &&  \mbox{\hspp}\left.
          - {805069021181 \over 254944609920}\*\zeta_4
          + {18769 \over 62424}\*\zeta_5
          \right)
\nn \\[0.5mm] && \mbox{\hspn}
       + \,\nf\, \*\cas\, \* \cf\, \* \left(
            {108178133162924948555733173122528196346341 \over 64633896443253521100663132819849216000000}
          + {12453233313645527413979 \over 7816915474984167552000}\*\zeta_3
\right.\nn \\[0.5mm] &&  \mbox{\hspp}\left.
          + {19557316769 \over 8852243400}\*\zeta_4
          - {131747 \over 46818}\*\zeta_5
          \right)
\nn \\[0.5mm] && \mbox{\hspn}
       + \,\nf\, \*\dfRRnr\, \* \left(
            {194754027746301317663486903 \over 2385646868143957017600000}
          + {488331702547711013 \over 28175156700490800}\*\zeta_3
          - {808523 \over 7803}\*\zeta_5
          \right)
\nn \\[0.5mm] && \mbox{\hspn}
       + \,\nfs\, \*\cfs\, \* \left(
          - {27940032477586559318231174397709681 \over 58566283098537309493828740710400000}
          - {220452248921 \over 382416914880}\*\zeta_3
          + {18769 \over 62424}\*\zeta_4
          \right)
\nn \\[0.5mm] && \mbox{\hspn}
       + \,\nfs\, \*\ca\, \*\cf \* \left(
            {5139790280893055367044834414063191 \over 7912907734820757808449287577600000}
          + {1631966041 \over 5730523200}\*\zeta_3
          - {18769 \over 62424}\*\zeta_4
          \right)
\nn \\[0.5mm] && \mbox{\hspn}
       + \,\nft\, \*\cf\, \* \left(
          - {13608819731912112034987483 \over 206666738093821415406796800}
          + {18769 \over 468180}\*\zeta_3
          \right)
%
%%;
%%STOP
\; , \\[2mm]
\label{eq:GpsN18}
 \lefteqn{ \hspn \gamma_{\,\rm ps}^{\,(3)}(N\!=\!18) \:\equal\:
%%START
%%L %%texGpsN18 =
%
       \nf \*\cft \* \left(
            {7253359571892497953990576741964828731260622709 \over 724111130666071269938038292783254417075200000}
\right.} \nn \\[0.5mm] &&  \mbox{\hspp}\left.
          - {109721806416706447798 \over 105229951281929298825}\*\zeta_3
          + {9447656272424 \over 12054216999825}\*\zeta_4
          - {236672 \over 165699}\*\zeta_5
          \right)
\nn \\[0.5mm] && \mbox{\hspn}
       + \,\nf\, \* \ca\, \*\cfs\, \*\left(
          - {28959853818114889791771647893787915791563566723 \over 4685424963133402334893188953303410934016000000}
\right.\nn \\[0.5mm] &&  \mbox{\hspp}\left.
          - {751673700996836590582 \over 526149756409646494125}\*\zeta_3
          - {30932168351824 \over 12054216999825}\*\zeta_4
          + {118336 \over 497097}\*\zeta_5
          \right)
\nn \\[0.5mm] && \mbox{\hspn}
       + \,\nf\, \* \cas\,\*\cf\, \*\left(
            {89162440521267641872038820315362221737605493 \over 45935538854249042498952832875523636608000000}
\right.\nn \\[0.5mm] &&  \mbox{\hspp}\left.
          + {132300696845690213098985759 \over 67414515989255185999248000}\*\zeta_3
          + {50551793128 \over 28362863529}\*\zeta_4
          - {522368 \over 165699}\*\zeta_5
          \right)
\nn \\[0.5mm] && \mbox{\hspn}
       + \,\nf\, \*\dfRRnr\, \* \left(
            {269658033224423391160920092267 \over 3660178707449866054252800000}
          + {231728400575045385953 \over 9154108411989460920}\*\zeta_3
\right.\nn \\[0.5mm] &&  \mbox{\hspp}\left.
          - {17270272 \over 165699}\*\zeta_5
          \right)
\nn \\[0.5mm] && \mbox{\hspn}
       + \,\nfs\, \* \cfs\,\*\left(
          - {510320588931393559068451118322848217263 \over 1393078636581692811620875836456225600000}
          - {10059181640656 \over 21697590599685}\*\zeta_3
          + {118336 \over 497097}\*\zeta_4
          \right)
\nn \\[0.5mm] && \mbox{\hspn}
       + \,\nfs\,\*\ca\,\*\cf\, \* \left(
            {5746360771534685866990135393992059327 \over 10978384977283253229773296769145600000}
          + {24382094512 \over 108379573425}\*\zeta_3
          - {118336 \over 497097}\*\zeta_4
          \right)
\nn \\[0.5mm] && \mbox{\hspn}
       + \,\nft\, \*\cf\, \* \left(
          - {12093315480521279173973861137 \over 220470747728360428912290678000}
          + {236672 \over 7456455}\*\zeta_3
          \right)
%
%%;
%%STOP
\; , \\[2mm]
\label{eq:GpsN20}
 \lefteqn{ \hspn \gamma_{\,\rm ps}^{\,(3)}(N\!=\!20) \:\equal\:
%%START
%%L %%texGpsN20 =
%
         \nf\, \*\cft\, \* \left(
            {2128032487727689123396891103081423002879945894061 \over 236298858959429600796016734112798923362304000000}
\right.} \nn \\[0.5mm] &&  \mbox{\hspp}\left.
          - {7463032385600125416449 \over 7804464223042296810000}\*\zeta_3
          + {9834028074797 \over 14900178793500}\*\zeta_4
          - {178084 \over 153615}\*\zeta_5
          \right)
\nn \\[0.5mm] && \mbox{\hspn}
       + \,\nf\, \*\ca\, \*\cfs\, \* \left(
          - {8442281731349030891500282315883757515259615913 \over 1413270687556397133947468505459323704320000000}
\right.\nn \\[0.5mm] &&  \mbox{\hspp}\left.
          - {13512345934144930064021 \over 10405952297389729080000}\*\zeta_3
          - {7936779238702 \over 3725044698375}\*\zeta_4
          + {89042 \over 460845}\*\zeta_5
          \right)
\nn \\[0.5mm] && \mbox{\hspn}
       + \,\nf\, \*\cas\, \*\cf\, \*\left(
            {250450109018215553669333751863263807123028219 \over 119012268425801863911365768880785154048000000}
\right.\nn \\[0.5mm] &&  \mbox{\hspp}\left.
          + {40625424437896114995230699 \over 18397723661785041013440000}\*\zeta_3
          + {164760066767 \over 112031419500}\*\zeta_4
          - {4694036 \over 1382535}\*\zeta_5
          \right)
\nn \\[0.5mm] && \mbox{\hspn}
       + \,\nf\, \*\dfRRnr\, \* \left(
            {124046988016629781809318499469746921 \over 1840118243383345660115052672000000}
          + {34660205433264885994007 \over 1100342324269440252000}\*\zeta_3
\right.\nn \\[0.5mm] &&  \mbox{\hspp}\left.
          - {48237328 \over 460845}\*\zeta_5
          \right)
\nn \\[0.5mm] && \mbox{\hspn}
       + \,\nfs\, \* \cfs\, \*\left(
          - {20553091730130297702276618606953655791 \over 71772053747957053386934630643251200000}
          - {2842660003013 \over 7450089396750}\*\zeta_3
          + {89042 \over 460845}\*\zeta_4
          \right)
\nn \\[0.5mm] && \mbox{\hspn}
       + \,\nfs\, \*\ca\, \*\cf\, \* \left(
            {688560020231378646396927215051130832957 \over 1602729320537086079392450022635008000000}
          + {1316792611 \over 7223745375}\*\zeta_3
          - {89042 \over 460845}\*\zeta_4
          \right)
\nn \\[0.5mm] && \mbox{\hspn}
       + \,\nft\, \*\cf\, \* \left(
          - {46235817346069201871585241841 \over 990993385042051234188945600000}
          + {178084 \over 6912675}\*\zeta_3
          \right)
%
%%;
%%STOP
\; .
\eea
}}
The results for $\gamma_{\,\rm ps}^{\,(3)}$ at $N \leq 8$ in 
eqs.~(\ref{eq:GpsN2}) -- (\ref{eq:GpsN8}) have been published before in 
ref.~\cite{Moch:2021qrk}, and 
the values at $N \leq 16$ of the terms proportional to the quartic group
invariant $\dfRRnR$ in eqs.~(\ref{eq:GpsN2}) -- (\ref{eq:GpsN16}) 
have already been obtained in ref.~\cite{Moch:2018wjh}.
The $\nft$ contributions are known at all $N$ \cite{Davies:2016ruz}.

A {\sc Form} file with our results for $\gamma_{\,\rm ps}^{}(N)$ at 
even $N \leq 20$ and all partial all-$N$ expressions in the main text, 
and a {\sc Fortran} subroutine of our approximations for the splitting 
function $P_{\,\rm ps}^{\,(3)}(x)$ can be obtained from the
preprint server {\tt http://arXiv.org} by downloading the source.
Furthermore they are available from the authors upon request.

\end{document}